\documentstyle[aps,prb,epsf,twocolumn,floats]{revtex}
\begin{document}
\draft
\twocolumn[\hsize\textwidth\columnwidth\hsize\csname @twocolumnfalse\endcsname
\title{Magnetoresistance of the double-tunnel-junction
Coulomb Blockade with magnetic metals}
\author{Kingshuk Majumdar and Selman Hershfield}
\address{Department of Physics and National High Magnetic
Field Laboratory, University of Florida,\\
215 Williamson Hall, Gainesville, FL 32611}
\date{\today}
\maketitle

\begin{abstract}
We have studied the Junction Magnetoresistance (JMR) and the 
Differential junction magnetoresistance (DJMR) for 
double tunnel junctions with magnetic metals in the Coulomb Blockade
regime.  Spikes are
seen in both the JMR and the DJMR vs. voltage curves. They 
occur at those places where the current increases by a step. 
In all cases the large bias limit can be obtained by adding the
resistances of each of the junctions in series.
 The JMR is positive in all the 
cases we studied, whereas the DJMR can be positive or negative as a function
of the voltage.
Moreover, the relative variation of the 
DJMR as a function of the voltage is larger than the variation
of the JMR with the voltage. 
\end{abstract}
\pacs{PACS numbers: 73.40.Gk, 73.23.Hk, 75.70.Pa}
\bigskip
]
\narrowtext
\section{INTRODUCTION}
Electron tunneling through tunnel junctions which consist
of magnetic metals separated by insulators has a long history.
\cite{wolf}
Some of the earliest experiments were on
tunneling between a superconductor and a ferromagnet.\cite{tedrow}
In these experiments, the difference 
between the spin-up and spin-down density of states of a
superconductor in a magnetic field was observed.  These experiments also provided 
a quantitative measure of the spin polarization of the current from the 
ferromagnetic metal.
Later experiments were done on tunneling
between two ferromagnetic metals.\cite{julliere}  Because of the
spin dependence of the density of states, the conductance changed when
the relative orientation of the two magnetizations changed.
Tunneling between two ferromagnetic metals has experienced renewed interest
because of its potential application in information storage. One of 
the measurable quantities of practical importance is the 
junction magnetoresistance, which is defined as 
the ratio of the change in the conductance from parallel to
antiparallel alignment divided by the parallel conductance. 
Large magnetoresistance response in low magnetic fields has been 
observed in a variety of tunnel junctions by several groups.
\cite{{tan},{plat},{mk},{nk},{gd},{naka},{ps},{clair},{miya2},{maki},{mood}}

Besides spin, electrons carry charge and at low temperatures in tunnel 
junctions with small capacitance, the discreteness of electron charge 
manifests itself through the physics of the Coulomb Blockade.
\cite{{gort},{kulik},{likh},{av},{schon},{grab}}
The transfer of an electron by tunneling between two initially
neutral regions of capacitance $C$ increases the electrostatic
energy of the system by an amount $e^2/2C$. At low temperatures and
small voltages, the tunneling current is suppressed because
of the charging energy. This energy barrier is called the Coulomb 
Blockade. The effect of this in a two-junction system is the incremental
increase in the current at voltages where it is energetically favorable
for an electron to tunnel to the center island. The occurrence of these current
steps in the current-voltage characteristics is known as the ``Coulomb 
Staircase.''
Advances in technology have led to a number of new experiments 
which exhibit the Coulomb Blockade phenomenon: granular thin              
films\cite{{raven},{hanna}},
lithographically patterned tunnel junctions\cite{{fulton},{ralph}},
narrow disordered quantum wires\cite{{chandra},{greg}},
scanning tunneling microscopy of small metal droplets\cite{wilkins}, 
metal-insulator-metal tunnel junctions with small metal 
particles embedded in the insulator.
\cite{{barner},{kuz},{bentum},{been}}

Recently there have been experiments where spin polarized tunneling 
and the Coulomb Blockade have been combined.\cite{sankar} 
For example, in discontinuous metal/insulator multilayers consisting of
closely spaced ferromagnetic nanoparticles
in an insulating matrix, a Coulomb Blockade is seen in the 
current-voltage characteristic at low temperatures.
The purpose of this paper is to further study  
the nonlinear conductances for parallel and antiparallel 
magnetization alignment for a 
double junction with magnetic metals in the Coulomb Blockade regime.

The organization of the rest of the paper is as follows. In the next 
section we describe the model and the master equation used. The expression 
for the current, the conductance, the differential conductance and 
the junction magnetoresistance are then introduced. The results of the 
calculation are given in the discussion of Sec. III, and summarized 
in Sec. IV.

\section{Formalism}
The experimental realization of a double tunnel junction 
consists of three metals separated by oxide layers which act as 
insulators. The two end leads, denoted by left ($L$) and
right ($R$), are connected to a constant voltage source $V$.
This double junction can be modeled as a closed loop electrical circuit with 
two leaky capacitors of capacitances $C_{\rm L}$ and $C_{\rm R}$ connected
in series (Fig. 1). The resistances of the junctions, $R_L$ and $R_R$, are 
connected in parallel to $C_L$ and $C_R$ respectively.
\begin{figure}[tb]
\protect \centerline{\epsfxsize=2.5in \epsfbox {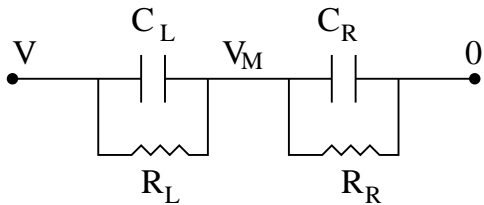}}
\vskip .2cm
\protect \caption{Schematic diagram of a double tunnel junction system.
The left and right capacitors ($C_L$ and $C_R$) are connected in series
with a voltage source.
The resistances of the left and right junctions are $R_L$ and $R_R$, 
respectively. The voltage in the left  and right leads are kept
fixed at $V$ and 0, whereas in the central region the voltage $V_M$ 
changes.}
\protect \label{fig1}
\end{figure}
Depending on the number of excess electrons, $n$,
 in the center metal, the voltage in the central region $V_M$ 
fluctuates. The voltage 
drops, $(V_L-V_M(n))$ and $(V_M(n)-V_R)$, across the left and the 
right junctions are    
\begin{eqnarray}
V_L-V_M(n) &=& \frac{C_R}{C}V+\frac{ne}{C}+V_G,\\
V_M(n)-V_R &=& \frac{C_L}{C}V-\frac{ne}{C}-V_G,
\end{eqnarray}
where $-e$ is the charge of the electron, $C$ is the capacitance equal to 
$(C_L+C_R)$, and $V_G$ plays the role of
the gate or the offset voltage.\cite{{lambe},{amman}} 
To describe the transport of electrons we use a semiclassical model.
\cite{{kulik},{amman},{selman},{lk}}
The different tunneling rates that enter
into the dynamics are: the rate for electrons to tunnel onto the 
center metal from the left ($\Gamma^L_{n\rightarrow n+1}$) and 
right ($\Gamma^R_{n\rightarrow n+1}$), and the rate for the 
electrons to leave the center metal to the left 
($\Gamma^L_{n\rightarrow n-1}$) and right 
($\Gamma^R_{n\rightarrow n-1}$). These tunnel rates can be calculated 
using Fermi's Golden rule.\cite{amman} To express the different 
tunnel rates in a convenient fashion we introduce a function 
$\gamma (\epsilon)$:
\begin{equation}
\gamma(\epsilon)=\frac{\epsilon}{1-e^{-\beta \epsilon}}.
\label{gamma}
\end{equation}
In terms of the above function the various tunnel rates are:
\begin{equation}
\Gamma^{L(R)}_{n \rightarrow n \pm 1}=\left(\frac 1{e^2R_{L(R)}}\right)
\gamma\left\{ \pm e\left[V_M(n)-V_{L(R)}\right]-E_C\right \}.
\end{equation}
Here $E_C=e^2/2C$ is the charging energy, and $\beta$ is related to
the temperature $T$ via $\beta=1/k_B T$.

For magnetic materials, there are two types of 
carriers: the majority electrons with spins parallel to the magnetization
and the minority electrons with spins antiparallel to the magnetization.
We make the assumption that 
the rate of tunneling through the junctions is much smaller than the rate 
of spin relaxation. Under this assumption the number of majority and 
minority electrons in the center metal is not conserved separately, but 
their sum is conserved. 
This results in a master equation which describes the time evolution of
the probability, $\rho_n(t)$, for $n$ excess 
electrons in the center metal\cite{kulik}   
\begin{eqnarray}
\frac{d\rho_n(t)}{dt} &=& \left[ \rho_{n+1}\Gamma_{n+1 \rightarrow n}-
\rho_n \Gamma_{n \rightarrow n+1}\right]	\nonumber \\
&-&\left[ \rho_{n}\Gamma_{n \rightarrow n-1}-
\rho_{n-1} \Gamma_{n-1 \rightarrow n}\right],
\label{master}
\end{eqnarray}
where the net tunnel rate $\Gamma_{i\rightarrow j}$ is equal to
$\left(\Gamma^L_{i\rightarrow j}+\Gamma^R_{i \rightarrow j}\right)$.  
At steady state, $d\rho_n (t)/dt=0$, the current in the left junction
equals the current in the right junction due to current conservation, 
and is given by 
\begin{equation}
I= e\sum_{n=-\infty}^{n=\infty}\rho_n \left(\Gamma^L_{n \rightarrow n-1}-
\Gamma^L_{n \rightarrow n+1}\right).
\label{current}
\end{equation}

To evaluate the current in general, Eqs. (\ref{master}) - (\ref{current}) 
have to be solved numerically. 
We follow the same procedure described in Ref. [35]. 
At low temperature and voltage, the higher charge
states $|n| >> 1$ are energetically forbidden and therefore we can
neglect them. 
We found that for our
choice of parameters it is 
sufficient to keep the 19 states around $n=0$. The highest and the lowest
states are $n=9$ and $n=-9$. 
The total conductance $G$  is then simply given by dividing the current with 
the voltage, $G=I/V$, and the differential conductance $G_{\rm diff}$ is 
obtained by differentiating the current with respect to the voltage,
$G_{\rm diff}=dI/dV$.

There are two possible choice of materials which lead to non-zero
magnetoresistance in a two junction system:
(1) all the leads are ferromagnets,
i.e. FM/I/FM/I/FM, and (2) the left or
the right lead is a paramagnet, i.e. FM/I/FM/I/PM or 
PM/I/FM/I/FM. Note that with the center lead non-magnetic 
(FM/I/PM/I/FM) there is no JMR or DJMR because the antiparallel 
conductance is identical to the parallel conductance within our 
assumption.
For each of these cases,
we calculate the conductances for 
two different relative orientations of the ferromagnet magnetizations:
parallel and antiparallel. Here
parallel (P) means that the magnetizations of all 
ferromagnets is in the same direction, and antiparallel (AP)
means the magnetization of the center ferromagnetic metal
has been reversed.  In case (1) of three ferromagnetic
metals, other configurations are also possible: the magnetization of the 
right or the left lead can be reversed keeping the magnetizations of the 
other two parallel. For case (2) where one of the left or the right lead
is non-magnetic, the magnetization of right or the left ferromagnetic lead
can also be reversed. However, the
qualitative features of the current, conductance, and the magnetoresistance
curves remain the same.
The junction magnetoresistance (JMR) is then defined as the ratio of the
change in the conductance from parallel, $G^{\rm P}$ to antiparallel 
alignments, $G^{\rm AP}$, divided by the parallel conductance
\begin{equation}
	{\rm JMR}=1-\frac{G^{\rm AP}}{G^{\rm P}}.
\label{jmr}
\end{equation}
The differential junction magnetoresistance (DJMR) is defined in the
same fashion as the JMR but with the total conductances replaced by 
the differential conductances, $G_{\rm diff}$, 
\begin{equation}
	{\rm DJMR}=1-\frac{G^{\rm AP}_{\rm diff}}{G^{\rm P}_{\rm diff}}.
\label{djmr}
\end{equation}

\section{DISCUSSION}
The different parameters of our problem are: the capacitances $C_L$ 
and $C_R$, the parallel resistances $R_L^{\rm P}$ and $R_R^{\rm P}$, 
the antiparallel resistances $R_L^{\rm AP}$ and $R_R^{\rm AP}$, the 
temperature $T$, and the gate or the offset voltage $V_G$. To estimate
the antiparallel resistances, we use 
Julliere's model for FM/I/FM tunnel junctions, where the 
junction magnetoresistance is expressed as a product of the magnitudes
of the polarizations of the ferromagnets:
\cite{julliere}
\begin{equation}
\frac{\triangle R}{R^{\rm AP}}= \frac {2P_1 P_2}{(1+P_1 P_2)}.
\label{jul}
\end{equation}
In the above equation, $\triangle R$ is the change in resistance
from antiparallel to parallel orientations, $R^{\rm AP}$ is
the junction resistance when the magnetizations of the ferromagnets 
are antiparallel, and $P_1$ and $P_2$ are the spin polarizations 
of the two FMs.

In Fig. (2) the currents and the total conductances are plotted for
parallel and antiparallel orientations of the magnetizations. 
Asymmetric junctions and low temperature are used to obtain the steps
in the current. 
Also, we take the offset or the gate voltage to be zero, $V_G=0$.
The effect of finite gate voltage is only to shift the 
$I-V$ curves.\cite{selman}
As expected in the Coulomb Blockade regime, the current increases by 
steps (case a), and the conductance shows oscillations with decreasing
amplitude (case b). Finally, for large enough voltage, the current 
becomes linear with voltage and the conductance approaches the 
classical limiting value which is the inverse of the resistance of 
two resistors connected in series
\begin{equation}
 G=\frac1{(R_L+R_R)}.
\end{equation}
\begin{figure}[tb]
\protect \centerline{\epsfxsize=3.6in \epsfbox {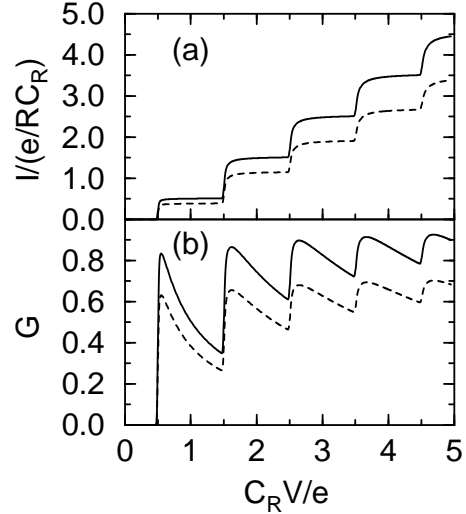}}
\vskip .2cm
\protect \caption{Current (I) and total conductance (G) as a function
of voltage, for a
FM/I/FM/I/FM tunnel junction. For both the cases (a) and (b), the solid
lines represent the current and the conductance for parallel alignment 
of the magnetizations, whereas the dashed lines are for antiparallel 
alignment. As a function 
of voltage the current shows steps (case a), and the total conductance 
oscillates with decreasing amplitudes (case b). 
We choose the left and the center metals to be made of iron with
polarization, $P_{\rm Fe}$=0.40, the right metal to be made of cobalt with 
polarization $P_{\rm Co}$=0.34.\cite{tedrow} 
The different parameters used are: $C_{\rm L}$=0.01$C$, 
$R_{\rm L}^{\rm P}$=0.01$(R_L^{\rm P}+R_R^{\rm P})$, and $k_B T=0.01E_C$.}
\protect \label{fig2}
\end{figure}

\begin{figure}[tb]
\protect \centerline{\epsfxsize=3.6in \epsfbox {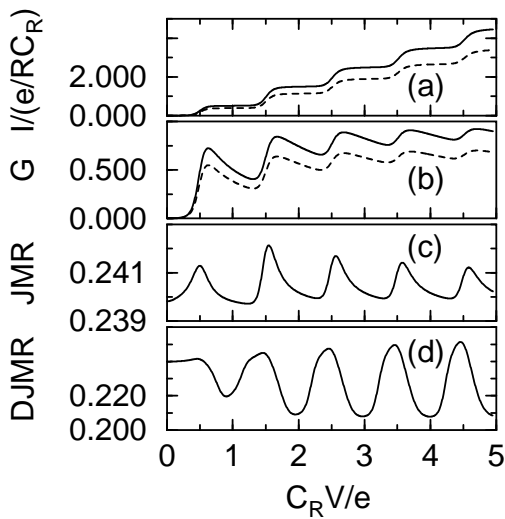}}
\vskip .2cm
\protect \caption{Temperature dependence of the (a) Current, 
(b) Conductances, (c) JMR, and (d) DJMR for a FM/I/FM/I/FM tunnel 
junction. As in Fig. 2(b), the 
solid and the dashed lines represent the current and the conductance 
for parallel and  antiparallel alignments. 
With increase in temperature, the steps in the current vs. 
voltage curve round off (case a), resulting in reduced amplitude of 
oscillations in the total conductance (case b). 
The peaks in both the JMR (case c) and the DJMR (case d) curves broaden.
Also the peaks in the JMR and the DJMR curves decreases with increase in
temperature.
The parameters are the same as used in Fig. 2, except for the 
temperature which is $k_B T=0.1E_C$.}
\protect \label{fig3}
\end{figure}

The effect of temperature dependence is shown in Fig. (\ref{fig3}).
With the rise in temperature, the number of accessible states increases. 
The steps in the current round off (case a), and the amplitude of the 
oscillations in the conductance vs. voltage curve (case b) decreases. 
The peaks in the JMR (case c) and the DJMR vs. voltage 
curves (case d) become broader and reduced in size.

Restricting ourselves to $k_B T << E_C$ and zero gate or offset 
voltage ($V_G=0$), in Fig. 4 we plot
the junction magnetoresistance for four different cases: 
(a) FM$_1$/I/FM$_1$/I/FM$_2$, 
(b) FM$_2$/I/FM$_1$/I/FM$_1$, 
(c) FM$_1$/I/FM$_1$/I/PM, and 
(d) PM/I/FM$_1$/I/FM$_1$, where FM$_1$ and FM$_2$ are different 
ferromagnetic metals.  
In the high bias regime for all four cases, the current is 
linearly proportional to the voltage\cite{selman}
\begin{equation}
I=\frac{e}{(R_L+R_R)C}\left[ \frac {CV}{e}-1\right].
\label{current3}
\end{equation}
It then follows from  Eq. (\ref{current3}), 
that the plateaus in Figs. 4(a) - 4(d) are given by the difference in the 
total resistances from antiparallel to parallel alignments divided by 
the total resistance for the antiparallel alignment, 
\begin{equation}
	{\rm JMR (plateau)}= 1-\frac{R_L^{\rm P}+R_R^{\rm P}}
			{R_L^{\rm AP}+R_R^{\rm AP}}.
\label{plateau}
\end{equation}

For asymmetric junctions in the low voltage, low temperature regime  
we see spikes in the JMR vs. voltage curves in Fig. (4).  
The occurrence of the first spike 
can be understood analytically. We confine ourselves to
only three states: the $n=0$ state and the two states $n=\pm 1$ immediately
accessible from it. At very low temperature, 
the state which is most likely occupied is the $n=0$ state for $V_G=0$.
The tunnel rates $\Gamma_{0 \rightarrow 1}$ and $\Gamma_{1 \rightarrow 0}$
do not enter in the expression for the tunneling current at zero temperature,
but at finite temperature they do contribute. 
In the limit of zero temperature, where we 
replace the function $\gamma (\epsilon)$ in Eq. (\ref{gamma}) by 
$\gamma(\epsilon)= \epsilon \theta (\epsilon)$, the steady state current
can be expressed in terms of the tunnel rates in the regime 
$(e/2) \leq C_R V \leq (C_R/C_L)(e/2)$ as
\begin{equation}
I=e \frac{\Gamma^L_{0 \rightarrow -1} \Gamma^R_{-1 \rightarrow 0}}
	{(\Gamma^L_{0 \rightarrow -1} + \Gamma^R_{-1 \rightarrow 0})}.
\label{current2}
\end{equation}
In the derivation of Eq. (\ref{current2}) we have assumed that the capacitance 
of the right junction is larger than the capacitance of the left junction,
i.e, $C_R > C_L$. 
Near the special point $C_R V \approx e/2$, where the 
current increases by a step at zero temperature, the height of the 
spike is simply given by the difference, 
\begin{equation}
	{\rm JMR}(C_R V=\frac{e}{2}, T\rightarrow 0)= 
		1-\frac{R^{\rm P}_L}{R^{\rm AP}_L}.
\label{height}
\end{equation}
However, at finite temperature, the height of the 
spikes depend on the temperature, the capacitances, and the 
resistances of the junctions. 
We also notice that spikes go up for cases 4(a) and 4(c) whereas spikes
go down for cases 4(b) and 4(d). 
Cases 4(a) - 4(b) or cases 4(c) - 4(d) differ only by the exchange of 
the metals in the left and the right leads. 
In all the above cases we keep the resistance of the left junction 
for the parallel alignment fixed, $R_L^{\rm P}=0.01(R_L^{\rm P} +
R_R^{\rm P})$. By only exchanging the metals of the left and the right 
leads, we go from spikes which go up to spikes which go down and vice versa. 
The interchange of the spikes from up to down or down to up can be explained
in the following way.  
At zero temperature, whenever the JMR obtained from Eq. (\ref{height})
is greater than the height of the plateau (Eq. (\ref{plateau})), we 
see an up-spike (Fig. 4(a) and 4(c)), otherwise we see a down-spike 
(Figs. 4(b) and 4(d)). Within Julliere's model\cite{julliere},
we should mention that to see the spikes, the left or the
right metal should be of a different ferromagnetic material. When all 
three metals are of the same ferromagnetic materials, the steady state
solution $\rho_n^0 (t)$ is the same for parallel and antiparallel 
orientation of magnetizations since the solution depends only on the 
ratio of the left and the right resistances. 
\begin{figure}[tb]
\protect \centerline{\epsfxsize=3.6in \epsfbox {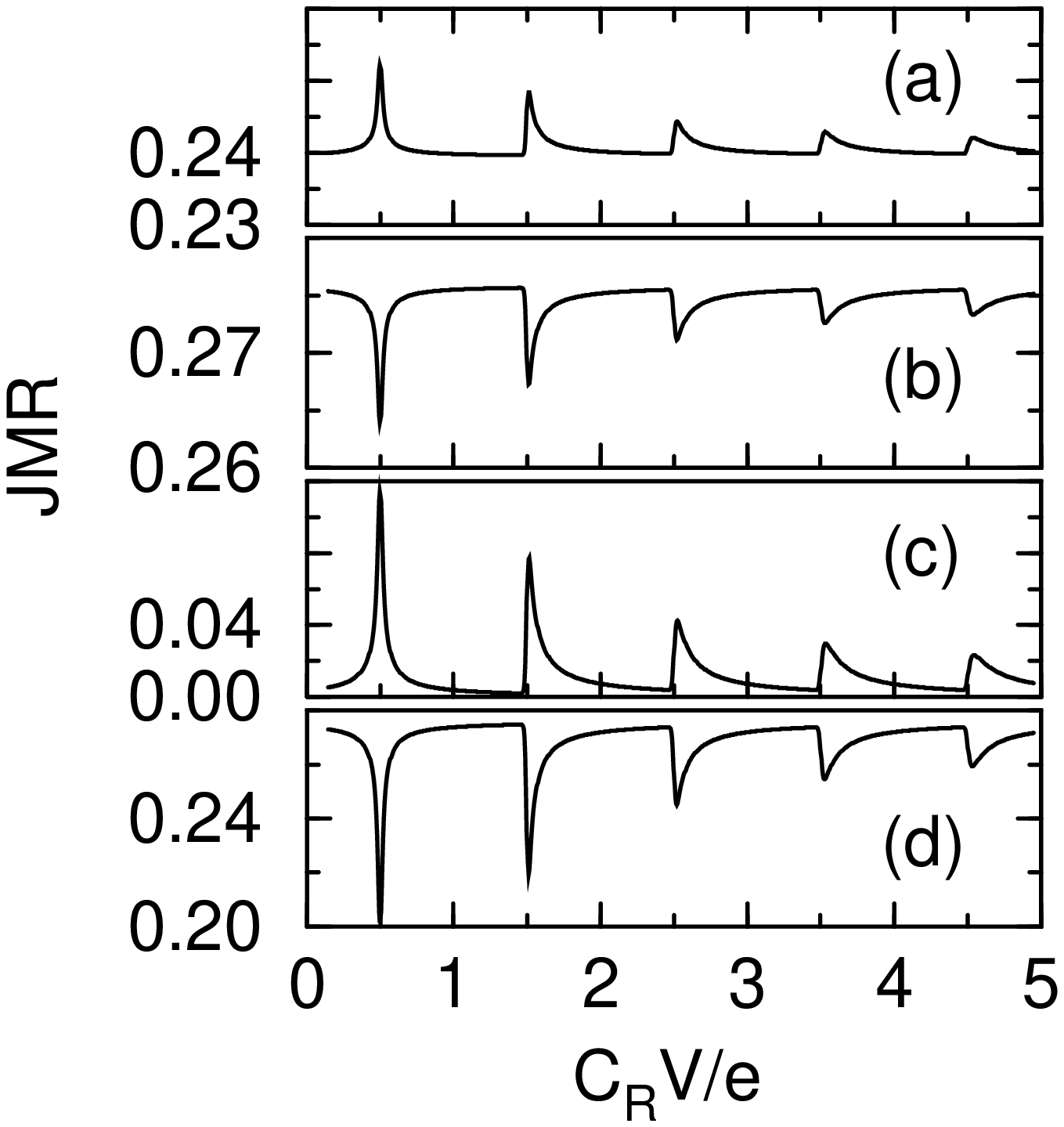}}
\vskip .2cm
\protect \caption{Junction magnetoresistance (JMR) as a function of 
voltage for different cases: 
(a) FM$_1$/I/FM$_1$/I/FM$_2$, 
(b) FM$_2$/I/FM$_1$/I/FM$_1$, 
(c) FM$_1$/I/FM$_1$/I/PM, and 
(d) PM/I/FM$_1$/I/FM$_1$, where FM$_1$ and FM$_2$ are two
different ferromagnets, and PM is any paramagnet. In all cases we
see spikes in the JMR: spikes go up for cases (a) and (c) whereas
spikes go down for cases (b) and (d).
These spikes occur at those places
where the current increases by a step (see Fig. 2(a)). 
The plateaus in all cases are given by the difference in the 
total resistances from antiparallel to parallel alignments divided 
by the total resistance for the antiparallel alignment.
As a function of voltage the variation of JMR is larger when one of the end
metals is a paramagnet (cases c and d).
For our plots we consider the ferromagnet FM$_1$ to be of iron, and 
FM$_2$ is of cobalt. The parameters chosen are the same as in Fig. 2.}
\protect \label{fig4}
\end{figure}

Comparing Figs. 4(a) - 4(b) with Figs. 4(c) - 4(d), we see that the 
variation of the JMR as a function of voltage is larger when one of 
the leads is non-magnetic. Following 
Eq. (\ref{plateau}) and Eq. (\ref{height}), which give the height 
of the plateau and the height of the first spike at zero temperature, we
see that the height of the plateau for case (c) is lower than for 
case (a).
On the other hand, the heights of the first spikes are same 
for both cases because the 
metals in the left and the center are same. Thus the variation
of the JMR for case (c) is more than case (a).
Cases (b) and (d) can be understood similarly.  
 
\begin{figure}[tb]
\protect \centerline{\epsfxsize=3.6in \epsfbox {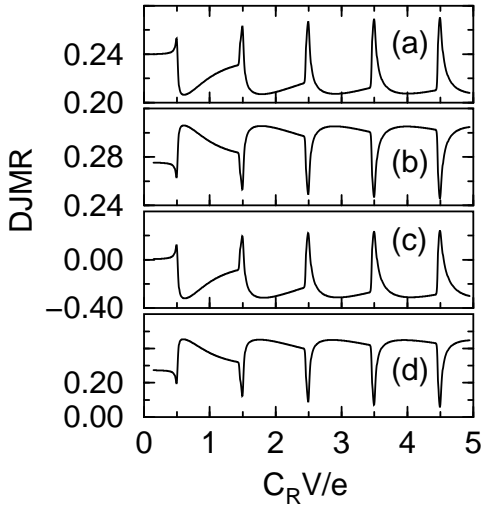}}
\vskip .2cm
\protect \caption{Differential Junction magnetoresistance (DJMR) 
as a function of voltage for the same cases as in Fig. 4. We
see spikes in the DJMR at those places where the current increases by 
a step.  As in Fig. 4, the variation in the DJMR is larger as a 
function of voltage for
cases (c) and (d) compared to cases (a) and (b). For case (c) the DJMR 
changes sign. Also compared to Fig. 4, the change in the DJMR is 
larger than the JMR. 
The metals are the same as used in Fig. 4 and the parameters are 
identical to Fig. 2.}
\protect \label{fig5}
\end{figure}

We plot the differential junction magnetoresistance as a function of
the voltage in Fig. 5 for the same cases as in Fig. 4 and in the same 
parameter regime.  
Many of the features of the DJMR are the same as for the JMR.
The spikes are at the same places. The height of the spikes is also 
the same as in Eq. (\ref{height}). The DJMR at large voltages is
obtained from the classical value with two resistors in series. 
Finally, the variation of the 
DJMR with the voltage is larger in cases 5(c) and 5(d) than in
cases 5(a) and 5(b). 
There are, however, some differences.
The DJMR can be negative, as shown in Fig. 5(c),
while the JMR must be positive from Eq. (\ref{height})
as long as $R_L^{\rm P} < R_L^{\rm AP}$ and $C_R > C_L$. Also,
comparing Fig. (\ref{fig4}) with Fig. (\ref{fig5}), we see that 
the variation of DJMR as a function of voltage is larger than the 
variation of the JMR with the voltage.

\section{CONCLUSION}
In this paper we have computed the junction magnetoresistance and 
the differential junction magnetoresistance 
for double tunnel junctions with magnetic metals in the Coulomb 
Blockade regime.
As expected in a Coulomb Blockade
problem, the steps in the current and the oscillations in the total 
conductance are observed for both the parallel and antiparallel 
orientations of the magnetizations, and the structure is reduced with 
increasing temperature. 

The JMR as a function of voltage shows spikes 
at the same places where the current increases by a step.
The height of the spikes in general depends on the 
temperature, capacitances, and the resistances of the system whereas
the height of the plateaus in the JMR vs. voltage curves are
given by the classical value obtained by adding two resistors in series.
For asymmetric junctions where $C_R > C_L$, the 
spikes go up if the 
ratio of the resistances of the right junction for parallel and 
antiparallel alignment ($R_R^{\rm P}/R_R^{\rm AP}$) is greater than 
the same ratio of the resistances of the left junction 
($R_L^{\rm P}/R_L^{\rm AP}$). Otherwise, the spikes go down.

Many of the features of the DJMR are similar to the JMR:
the spikes occur at the same places, the height of the first
spike at zero temperature is the same, and the sign (up or down) of the 
spikes are the 
same. Also at high bias, the DJMR saturates to the classical
value obtained by adding two resistors in series. However, there are few
differences. The DJMR can be negative for some cases, while the JMR is 
positive as long as $R_L^{\rm P}<R_L^{\rm AP}$ and $C_R > C_L$.
Finally, the variation of the 
DJMR is larger as a function of voltage than that of the JMR .

In conclusion, double tunnel junction systems which have magnetic
metals exhibit rich Coulomb blockade conductance vs. voltage curves.
These curves provide a signature of both 
the Coulomb charging effects and the
spin polarization of the tunneling.  By measuring the JMR or DJMR,
one can extract information about both the capacitance charging
energies and the spin polarization of the tunneling electrons,
testing both theories of spin polarized tunneling and the Coulomb
blockade.  Furthermore, there are cases where the JMR can be 
dramatically enhanced near one of the Coulomb blockade steps,
meaning that there may be some applications of these effects.
In any case, for small enough systems, both charging and spin-polarization
effects will be important.

\acknowledgments

This work was supported by DOD/AFOSR grant F49620-96-1-0026 and
NSF grant DMR9357474, and the NHMFL. We would like to
thank J. Chen for useful discussions.

\end{document}